\documentclass[twocolumn,,amsmath,amssymb]{revtex4}

\usepackage{graphicx}
\usepackage{dcolumn}
\usepackage{bm}
\newcommand{\etal}{\emph{et al.}}
\newcommand{\be}{\begin{equation}}
\newcommand{\ee}{\end{equation}}
\newcommand{\bfig}{\begin{figure}}
\newcommand{\efig}{\end{figure}}

\begin{document}
\title{Phase fluctuations versus Gaussian fluctuations in optimally-doped YBa$_2$Cu$_3$O$_7$.
}
\author{Lu Li$^{1,2}$, Yayu Wang$^{1,3}$ and N. P. Ong$^1$
}
\affiliation{
$^1$Department of Physics, Princeton University, Princeton, NJ 08544\\
$^2$Department of Physics, University of Michigan, Ann Arbor, MI 48109\\
$^3$Department of Physics, Tsinghua University, Beijing, China
}

\date{\today}

\begin{abstract}
We analyze recent torque measurements of the magnetization $M_d$ vs. magnetic field $H$ 
in optimally doped YBa$_2$Cu$_3$O$_{7-y}$ (OPT YBCO) to argue against a recent proposal
by Rey \etal~that the magnetization results above $T_c$ are consistent with Gaussian fluctuations.
We find that, despite its strong interlayer coupling, OPT YBCO displays an 
anomalous non-monotonic dependence of $M_d$ on $H$ which represents direct evidence for
the locking of the pair wavefunction phase $\theta_n$ at $T_c$ and the subsequent
unlocking by a relatively weak $H$. These unusual features characterize the unusual nature of the
transition to the Meissner state in cuprate superconductors. They are absent in 
low-$T_c$ superconductors to our knowledge. We also stress the importance of the vortex liquid
state, as well as the profiles of the melting field $H_m(T)$ and the upper critical field curve $H_{c2}(T)$
in the $T$-$H$ plane. Contrary to the claims of Rey~\etal, we show that the curves of the 
magnetization and the Nernst signal illustrate the inaccessibility 
of the $H_{c2}$ line near $T_c$. The prediction of the $H_{c2}$ line by Rey \etal~is shown to
be invalid in OPT YBCO.
\end{abstract}

\maketitle                   
\section{Introduction}\label{sec:intro}
In the mean-field Gaussian treatment of fluctuations (valid in low-$T_c$ superconductors), 
the pair-wavefunction amplitude $|\Psi|$ vanishes at the critical temperature $T_c$. Above $T_c$,
amplitude fluctuations about the equilibrium point $|\Psi|$ = 0 may be regarded (in Schmid's 
elegant depiction~\cite{Schmid}) as droplets of condensate
of radius $\xi_{GL}$, the Ginzburg-Landau (GL) coherence length. 
In the competing phase-disordering scenario, 
$|\Psi|$ remains finite above $T_c$. The collapse of the Meissner effect above $T_c$ is
caused instead by the vanishing of phase rigidity. 
Above $T_c$, fluctuations are
primarily of the phase $\theta$ of $\Psi$, proceeding by phase slips caused by
the motion of spontaneous vortices. In zero magnetic field, the net vorticity in the sample is zero, so the
populations of ``up'' and ``down'' vortices are equal.
In underdoped cuprates, there is strong evidence from Nernst~\cite{WangPRB}, magnetization~\cite{WangPRL,LiEPL,LiPRB}
and other experiments in support of the phase-disordering scenario.

Initially, it seemed to us that optimally-doped YBa$_2$Cu$_3$O$_{7-y}$ (OPT YBCO), which has the largest interlayer
coupling energy (and lowest electronic anisotropy) among the known cuprates, would be 
most amenable to standard mean-field (MF) treatment, i.e. its fluctuations above $T_c$ are strictly
Gaussian. Indeed, in the early 90s, several groups applied the conventional
Maki-Thompson Aslamazov-Larkin theory to analyze fluctuation conductivity above $T_c$~\cite{Semba}.
Subsequently, these mean-field ``fits'', seemingly reasonable in OPT YBCO, 
were found to be woefully inadequate in underdoped YBCO and in most other cuprates.
The steady accumulation of evidence from Nernst and torque experiments
favoring the phase-disordering mechanism in 
underdoped hole-doped cuprates and in OPT Bi$_2$Sr$_2$CaCu$_2$O$_{8+y}$ (Bi 2212) 
has prompted a re-assessment of the case for OPT YBCO. The recent torque magnetometry results in Ref.~\cite{LiPRB} have persuaded us that
OPT YBCO is, in fact, much closer to the other hole-doped cuprates. The collapse of its Meissner state
at $T_c$ is also caused by vanishing of phase rigidity.

In a Comment, Rey \etal~\cite{Rey} have fitted the magnetization curves for OPT YBCO in Ref. \cite{LiPRB} 
to their model and claimed that the diamagnetic signal is consistent with
MF Gaussian fluctuations in a layered superconductor. Furthermore, they have extended their
calculation to temperataures $T<T_c$ to infer an upper critical field $H_{c2}$ that rises linearly
with reduced temperature $(1-T/T_c)$ with a slope of -3 T/K.

Here, we show that the fluctuation signal above $T_c$ is 
just the tip of the iceberg. Because of strong interlayer coupling in OPT YBCO, it is
necessary to go below $T_c$ to uncover the close similarities with other cuprates.
In Sec. \ref{sec:nonmono} we describe the non-monotonic variation of the magnetization curve just below $T_c$, previously noted in
torque measurements on OPT Bi 2212 ~\cite{LiEPL}. This characteristic feature -- inherent in the 
phase-disordering scenerio -- is common to the cuprates investigated to date.
In Sec. \ref{sec:closure}, we discuss the upper critical field $H_{c2}$ for $T$ close $T_c$ in OPT YBCO. The striking inability of
high-resolution experiments to detect the $H_{c2}$ line, a corner stone of the mean-field Gaussian
picture, is also a characteristic feature of the phase disordering mechanism. Finally, in Sec. \ref{sec:fluc}
we discuss the fits to the magnetization above $T_c$. We show that the 
conclusions of Rey \etal~\cite{Rey} are not valid.

\begin{figure}[t]
\includegraphics[width=8 cm]{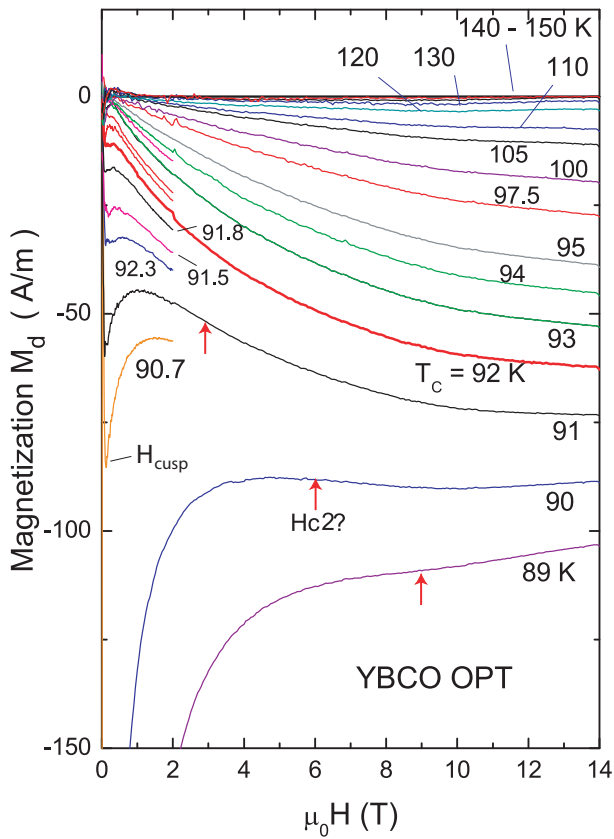}
\caption{\label{figMH} (color online) 
The field dependence of the diamagnetic component of the 
magnetization $M_d$ in optimally doped YBa$_2$Cu$_3$O$_{7-y}$ at selected temperatures $T$. 
At $T_c$ = 92.5, a weak step-increase in $|M_d|$ signals the onset of
full flux expulsion. Below $T_c$, a field $H>H_{cusp}$ causes 
a sharp decrease in the screening current, but at larger $H$, $|M_d|$
resumes its increase at a rate similar to that above $T_c$. This non-monotonic
pattern signals the field-induced decoupling of $\theta_n$
between adjacent bilayers below $T_c$. We argue that this is the 
defining magnetization feature that characterizes the transition in cuprates. 
$\mu_0$ is the vacuum permeability. The arrows (labelled as $H_{c2}?$) are
$H_{c2}$ values at 89, 90 and 91 K predicted in Ref. \cite{Rey}. (Adapted from Ref. \cite{LiPRB}.)
}
\end{figure}

\section{Non-monotonic magnetization} \label{sec:nonmono}
The curves of the diamagnetic component of the magnetization $M_d$ in OPT YBCO (Ref. \cite{LiPRB}) 
are reproduced in Fig \ref{figMH}. At temperature $T$ above $\sim$100 K, $M_d$ is initially linear in $H$
up to 6 T. As we approach $T_c$, the
linear-response regime becomes confined to progressively smaller field ranges (for e.g. $|H|<$ 2 T at 95 K).

At $T_c$ = 92.5 K (the present analysis shows that $T_c$ is slightly higher than the nominal value in Ref. \cite{LiPRB}), 
the only significant change is a step-increase in the magnitude $|M_d|$ near $H = 0^{+}$. Above 0.5 T, 
$M_d(H)$ is strikingly similar to the curves above $T_c$ apart from vertical scale. We
focus on the region in Fig. \ref{figMH}
bounded by the curves at $T_c$ and 90 K (the curve at 91 K is representative). 
At $H = 0^+$, the step-increase in $|M_d|$ signals the onset of full flux expulsion. 
However, a weak field interrupts this steep rise to produce a sharp cusp at the field $H_{cusp}$ ($H_{cusp}$
is slightly higher than the lower critical field $H_{c1}$ because of surface pinning of vortices). When $H$
exceeds $H_{cusp}$, $|M_d|$ falls rapidly to a broad minimum near 1 T, 
but subsequently \emph{rises} to even larger values. This non-monotonic profile in $M_d$, 
appearing just below $T_c$, is a defining hallmark of the transition in cuprates.
In this interval (90-92.5 K), $M_d$ vs. $H$ is reversible except in the
vicinity of $H_{cusp}$, where surface-barrier pinning of vortices causes a slight hysteresis.

\begin{figure}[t]
\includegraphics[width=9 cm]{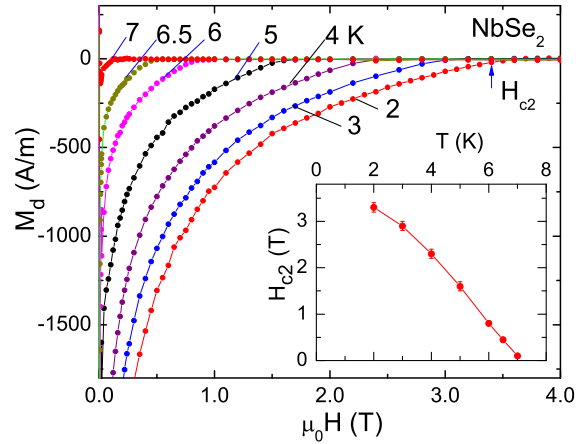}
\caption{\label{figNb} (color online) 
The magnetization curves below $T_c$ in NbSe$_2$. At each $T$,
$M_d(H)$ decrease monotonically to reach zero at $H_{c2}(T)$. Following
vortex entry at $H_{c1}$ (not resolved here), the diamagnetic screening currents
are steadily weakened as $H$ suppresses the pair amplitude. The inset plots the 
inferred $H_{c2}$ vs. $T$. (Adapted from
Ref. \cite{WangPRL})
}
\end{figure}

This unsual non-monotonic profile -- absent in low-$T_c$ superconductors
to our knowledge -- implies the following picture. The collapse of the Meissner state  
at $H_{cusp}$ leads to a steep decrease in $|M_d|$ as vortices enter the sample. This
is a consequence of field suppression of the nascent inter-bilayer phase coupling. 
(The two CuO$_2$ layers bracketing the Y ions constitute a bilayer. We are concerned 
with only the coupling between adjacent bilayers which are separated by a spacing $d$ = 11.8 \AA.
We mention the intra-bilayer coupling below.)
In low-$T_c$
superconductors, the decrease invariably continues monotonically to zero as $H\to H_{c2}(T)$
(the upper critical field at temperature $T$).
For reference, we show in Fig. \ref{figNb} curves of $M_d$ vs. $H$ in the
layered superconductor NbSe$_2$~\cite{WangPRL}.
By contrast, the non-monotonicity in $|M_d|$ implies that
the diamagnetic current in OPT YBCO turns around and grows \emph{stronger} with $H$ at a rate
closely similar to that above $T_c$.
Remarkably, if we hide the field region around $H_{cusp}$, 
the profiles of $M_d$ vs. $H$ below $T_c$ resemble 
those above $T_c$, apart from a different vertical scale.
This implies that the transition at $T_c$ at which 
flux expulsion appears is a weak-field phenomenon. Beyond this
weak-field regime, there is no hint that a transition has occured.

Returning to Fig. \ref{figMH}, we interpret the unusual non-monotonic pattern 
(now seen in Bi 2201, Bi 2212, LSCO and YBCO)~\cite{WangPRL,LiEPL,LiPRB}
as reflecting the rapid growth of $c$-axis phase rigidity below $T_c$ (in zero or weak $H$) 
and its subsequent destruction at $H_{cusp}$. 
In a finite interval above $T_c$, the pair wavefunction $|\Psi_n|{\rm e}^{i\theta_n}$ in bilayer $n$ 
has a finite average amplitude $\langle|\Psi_n|\rangle$. 
The in-plane phase-phase correlation length $\zeta_a$ in each bilayer, given by
\be
\langle {\rm e}^{-i\theta_n(0)}{\rm e}^{i\theta_n({\bf r})}\rangle = {\rm e}^{-r/\zeta_a},
\label{eq:corr}
\ee
is long enough that local diamagnetic currents can be detected by torque magnetometry. 
However, the $c$-axis
correlation length $\zeta_c\ll d$, so $\theta_n$ is
uncorrelated between adjacent bilayers. 
Hence $M_d$ reflects the diamagnetic response of 2D supercurrents which are observable to
very intense $H$. In Fig. \ref{figMH}, this 2D diamagnetic response is represented in 
the field profile of $M_d$ at 94 K.

Below $T_c$, $\zeta_c$ diverges (in zero $H$) 
to lock $\theta_n$ across all bilayers. In weak $H$, the 3D phase rigidity produces
full expulsion. However, in the 2-K interval 90-92.5 K, a weak field $H\sim H_{cusp}$ suffices
to destroy the $c$-axis phase stiffness. The system then 
reverts back to the diamagnetic response of 2D uncorrelated condensates. Consequently, as $H$
increases further, the 2D diamagnetic response continues to increase, mimicking the 
profile at 94 K. At 14 T, the 2D response leads to a value for $|M_d|$
that well exceeds its value at $H_{cusp}$ (see 91-K curve). The weak 
locking of $\theta_n$ across bilayers, and the field-induced unlocking, 
account for the non-monotonicity of $M_d$ as well as
the similarities of the high-field portions across $T_c$ in a physically 
reasonable way.
The juxtaposition of an extremely large pairing energy scale ($d$-wave gap amplitude
$\Delta\sim$40 mV) and a weak $c$-axis phase stiffness leads to this very unusual phase locking-unlocking
scenario which seems pervasive in the hole-doped cuprates. We argue that this
situation cannot be treated by applying
the mean-field GL approach~\cite{Schmid} to the Lawrence-Doniach (LD) Hamiltonian~\cite{Yamaji,Vidal94}.

\begin{figure}[t]
\includegraphics[width=8 cm]{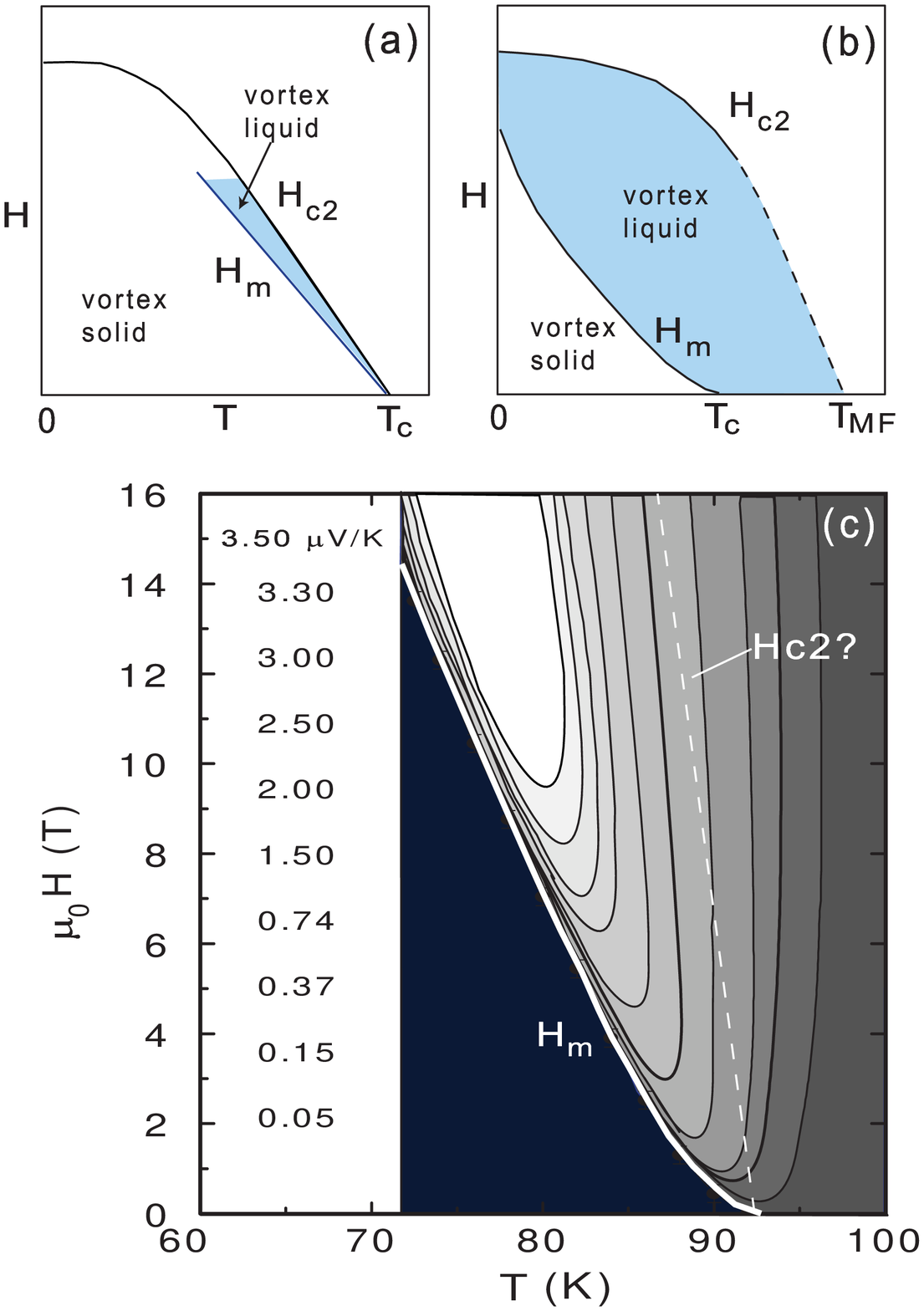}
\caption{\label{figphase} (color online) 
Profiles of the melting field $H_m(T)$ and upper critical field $H_{c2}(T)$ in low-$T_c$ and cuprate superconductors.
Panel (a) shows the phase diagram for a conventional type-II superconductor in the $T$-$H$ plane. The vortex liquid phase is wedged 
between the curves of $H_m(T)$ and $H_{c2}(T)$, both of which terminate at $T_c$ as $H\to 0$.
In the hole-doped cuprates [Panel (b)], the vortex liquid state dominates the phase diagram. As $H\to 0$, $H_m(T)$ terminates
at the observed $T_c$ whereas $H_{c2}(T)$ terminates at a higher temperature $T_{MF}$.
Panel (c): The contour plots of the magnitude of the Nernst signal $e_y$ in OPT YBCO in the 
$T$-$H$ plane. Values of the 10 contours are displayed on the left column. The melting field $H_m(T)$
separates the vortex solid (black region) from the vortex liquid state.
The dashed line is the $H_{c2}$ line predicted in Ref. \cite{Rey}. (Adapted from Ref. \cite{Wang2002}.)
}
\efig

\section{Closure of the $H_{c2}$ curve}\label{sec:closure}
Another strong argument against the mean-field description is obtained from 
the qualitative features of the phase diagram in the $T$-$H$ plane (Fig. \ref{figphase}).
A cornerstone of the mean-field Gaussian description of a type-II superconductor is 
the well-known profile of the upper critical field curve $H_{c2}(T)$ (Fig. \ref{figphase}a). In the $T$-$H$ plane,
the $H_{c2}$ curve is a sharply defined boundary that separates the region with finite pair amplitude $|\Psi|$ 
from the normal state with $|\Psi| = 0$. The Gaussian fluctuations are fluctuations of the amplitude 
around the $|\Psi|=0$ state above $H_{c2}(T)$. In low-$T_c$ superconductors (for $T>\sim 0.5\, T_c$), 
$H_{c2}(T)$ decreases linearly in the reduced temperature $t = 1-T/T_c$ to terminate at $T_c$. 
This profile is shown for NbSe$_2$ in the inset of Fig. \ref{figNb}.
To stress this point, we refer to the termination as ``closure'' of the curve of $H_{c2}$. 

The closure of $H_{c2}$ reflects an anomalous aspect of the BCS transition at $T_c$.
As $T$ is elevated above $T_c$ in zero $H$, the pair condensate vanishes before it loses its phase rigidity.
This implies that both the superfluidity (which depends on finite phase rigidity) and 
the physical entities that manifest long-range phase coherence (the Cooper pairs) vanish at the same temperature $T_c$.
(This is akin to having the local moments in an antiferromagnet vanish at the Ne\'{e}l temperature.)
In an applied field $H$, phase rigidity is maintained if the vortices are in the solid state (and pinned).
As $H$ approaches $H_{c2}$, the phase rigidity abruptly vanishes at the melting field $H_m$ which marks the 
transition to the vortex-liquid state (the analog of the paramagnetic state in the antiferromagnet). 
The vortex-liquid state is wedged between the curves of $H_m(T)$
and $H_{c2}(T)$ (Fig. \ref{figphase}a). As sketched, the anomalous feature in the BCS scenario is that, as $H\to 0$, 
the vortex-liquid region vanishes.

\begin{figure}[t]
\includegraphics[width=8 cm]{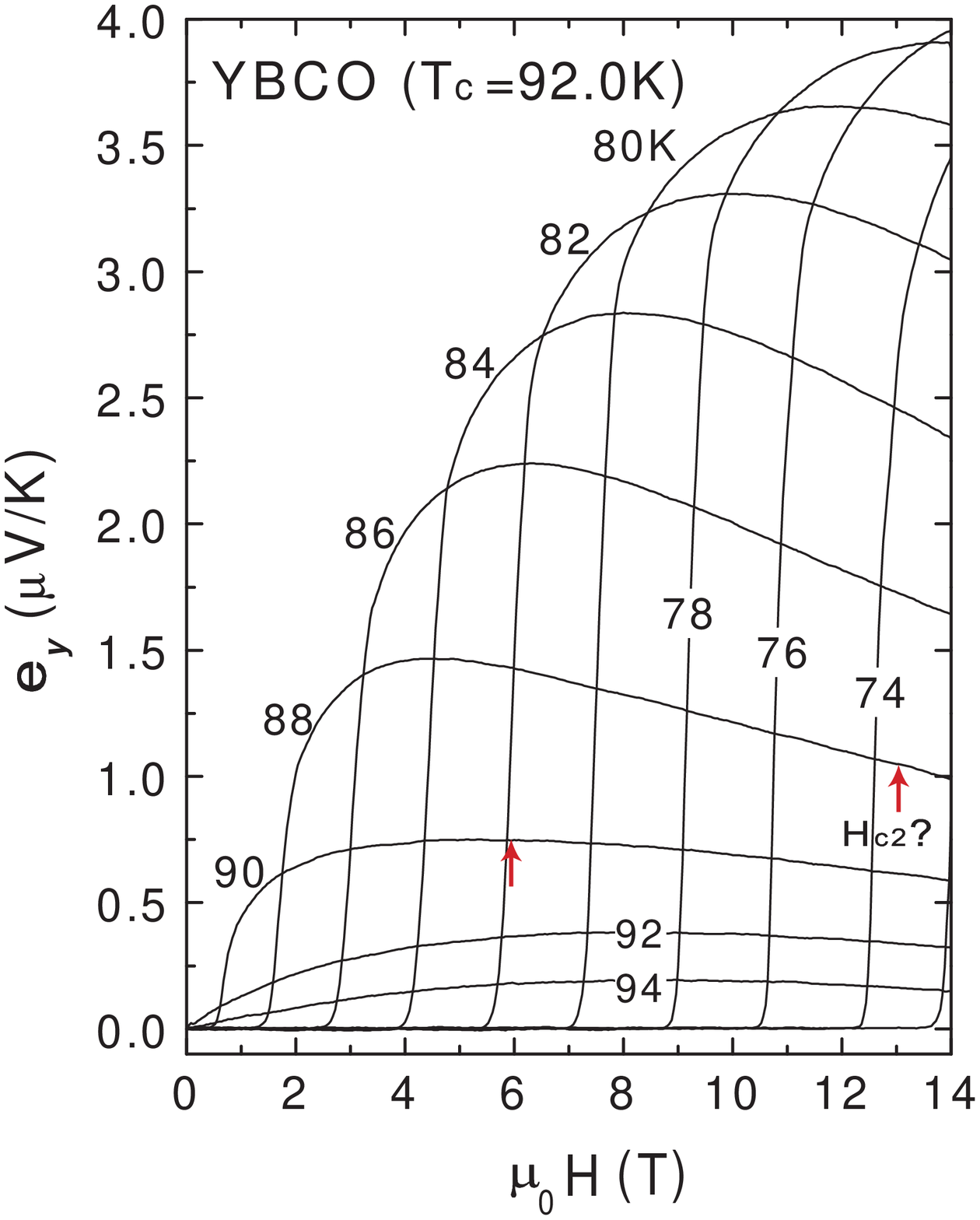}
\caption{\label{figNernst} (color online) 
Plots of the Nernst signal $e_y = E_y/|\nabla T|$ vs. $H$ in OPT YBCO with
$T_c$ = 92 K at selected $T$ ($E_y$ is the Nernst $E$-field
and $-\nabla T$ the applied temperature gradient). At the melting field $H_m$, $e_y$
rises nearly vertically reflecting the sharp increase in vortex velocity. The contour map (Fig. \ref{figphase}c)
was derived from these curves (and additional curves not plotted). The ``up'' arrows
denote the values for $H_{c2}$ at 88 and 90 K predicted in Ref.~\cite{Rey}. (Adapated from Ref. \cite{Ong2004}.)
}
\efig

The competing phase-disordering scenario has a qualtitatively different phase diagram (Fig. \ref{figphase}b).
The vortex liquid state now occupies a much larger 
fraction of the region in which $|\Psi|\ne 0$. Significantly, the melting curve $H_m(T)$ 
intercepts the $H=0$ axis at a temperature $T_m(0)$ lower than the termination point $T_{MF}$ of the
$H_{c2}$ curve, as sketched in Panel (b). If we increase $T$ along the $H=0$ axis, phase rigidity is lost at
$T_m(0)$ long before we reach $T_{MF}$. Since both the Meissner effect and the zero-resistivity state 
are crucially dependent on having long-range phase stiffness, the $T_c$ commonly observed by these techniques
is identified with $T_m(0)$. 

Above $T_c$, we have a vortex liquid whose existence 
may be detected using the Nernst effect and torque magnetometry (resistivity is ineffectual). 
The onset temperatures $T_{onset}$ for the vortex Nernst and diamagnetic signals~\cite{WangPRB,LiPRB} 
are lower bounds for $T_{MF}$.
Clearly, if the interval $[T_c,T_{MF}]$ is large (underdoped cuprates), the curve of $H_{c2}(T)$ is nearly
$T$ independent in the interval $0>T>T_c$; it attains closure only at $T_{MF}$. 
(The $T$-independent part has been
established in underdoped, single-layer Bi 2201 where $H_{c2}$ ($\sim$50 T) is nearly accessible with a 45-T magnet~\cite{LiPRB}.)
Hence, if we focus on experiments close to $T_c$, the only curve that intersects the $H = 0$ axis
is the melting curve. Even at $T_c$, $H_{c2}$ is inaccessible except for Bi 2201. 
The absence of closure for $H_{c2}(T)$ implies that the vortex-liquid state
extends above $T_c$ to $T_{MF}$, and survives to $H\sim H_{c2}$.

A striking empirical fact in the hole-doped cuprates is that experiments (over a 25-year period) have never observed an
$H_{c2}$ curve that decreases linearly with $t$ to terminate at $T_c$ (not counting early 
flux-flow resistivity experiments that mis-identified $H_m$ for $H_{c2}$). The absence of the $H_{c2}$ curve is
incompatible with the Gaussian picture, but anticipated in 
the phase-disordering scenario. 

We now turn to results in OPT YBCO. Figure \ref{figphase}c displays the contour plots of the
Nernst signal $e_y$ in another OPT YBCO crystal with identical $T_c$ and closely comparable quality 
(adapted from Ref. \cite{Wang2002}). The Nernst signal is defined as $e_y = E_y/|\nabla T|$, where
$E_y$ is the transverse electric field produced by the velocity of vortices diffusing in the applied thermal gradient $-\nabla T$ (see Ref. \cite{WangPRB}).
In the vortex solid ($H<H_m(T)$), $e_y$ is rigorously zero. Just above $H_m(T)$, $e_y$ rises very steeply
reflecting the steep increase in vortex velocity in the liquid state. In comparison with 
similar contour plots for underdoped cuprates (as well as OPT Bi 2212), the interval [$T_c$, $T_{MF}$] in OPT YBCO
is smaller. Also, the curve $H_m(T)$ rises from $T_c$ with a much steeper slope. 
Nevertheless, the contour features are qualitatively similar. The contours are very nearly vertical at $T_c$,
implying that $e_y$ hardly changes with $H$. Even exactly at $T_c$, there is no field scale above which the system 
can be described as being in the ``normal state'' with $|\Psi| = 0$. In cuprates, we find it helpful
to regard the segment of the $H=0$ axis between $T_c$ and $T_{MF}$ as the continuation of the $H_m(T)$ curve.
The vortex liquid, confined between the curves of $H_m$ and $H_{c2}$ below $T_c$, 
expands to occupy the entire region below $H_{c2}$ above $T_c$. It plays a dominant role in 
thermodynamic measurements.

In Ref. \cite{Rey}, the $H_{c2}$ line is
predicted to increase linearly with $t$ from the point ($T=T_c, H=0$) with 
a slope of -3 T/K. We have drawn their $H_{c2}$ line as a dashed line (labelled as $Hc2?$) in
Fig. \ref{figphase}c. 
Clearly, the predicted $H_{c2}$ line cuts across the contours in an arbitrary way that bears no resemblance
to experiment.

To explain this better, we show in 
Fig. \ref{figNernst} the profiles of $e_y$ vs. $H$ at selected fixed $T$
near $T_c$~\cite{Ong2004}. Just below $T_c$ (curves at 90 and 88 K), 
$e_y$ rises nearly vertically when $H$ exceeds $H_m$, reflecting the sharp increase
in the vortex velocity in the vortex liquid in response to the gradient $-\nabla T$ .
The magnitude of $e_y$, which remains large up to 14 T, implies that $|\Psi|$ remains finite.
There is no experimental feature (change in slope, e.g.) 
that signals $|\Psi|\to 0$ at the values of $H_{c2}$ predicted by Rey \etal ~(shown as ``up'' arrows).

The same difficulty exists for the plots of $M_d$ vs. $H$ in Fig. \ref{figMH}, where
the predicted $H_{c2}$ values are indicated by the arrows. Contrary to the claim in Ref. \cite{Rey},
the curves of $M_d$ vary smoothly through the predicted field values at 89, 90 and 91 K. 
Again, there is no feature reflecting either a transition or crossover.
Given that $|M_d|$ is a measure of the supercurrent
density $J_s$ surrounding each vortex in the sample (even in the vortex liquid state), we would expect $|M_d|$ to be considerably larger
when $|\Psi|$ is finite ($H<H_{c2}$) compared with the Gaussian fluctuation regime above $H_{c2}$. Instead,
$|M_d|$ hardly varies as $H$ crosses the predicted values.
One confronts the very awkward problem of explaining why $|M_d|$ retains nearly the same value below and above
``$H_{c2}$''. Even worse, close to $T_c$ (curve 91 K), $|M_d|$ actually
increases when $H$ exceeds 3 T (this reflects the decoupling of the bilayers as discussed in Sec. \ref{sec:nonmono}).

\section{Fluctuations above $T_c$}\label{sec:fluc}

\begin{figure}[t]
\includegraphics[width=8 cm]{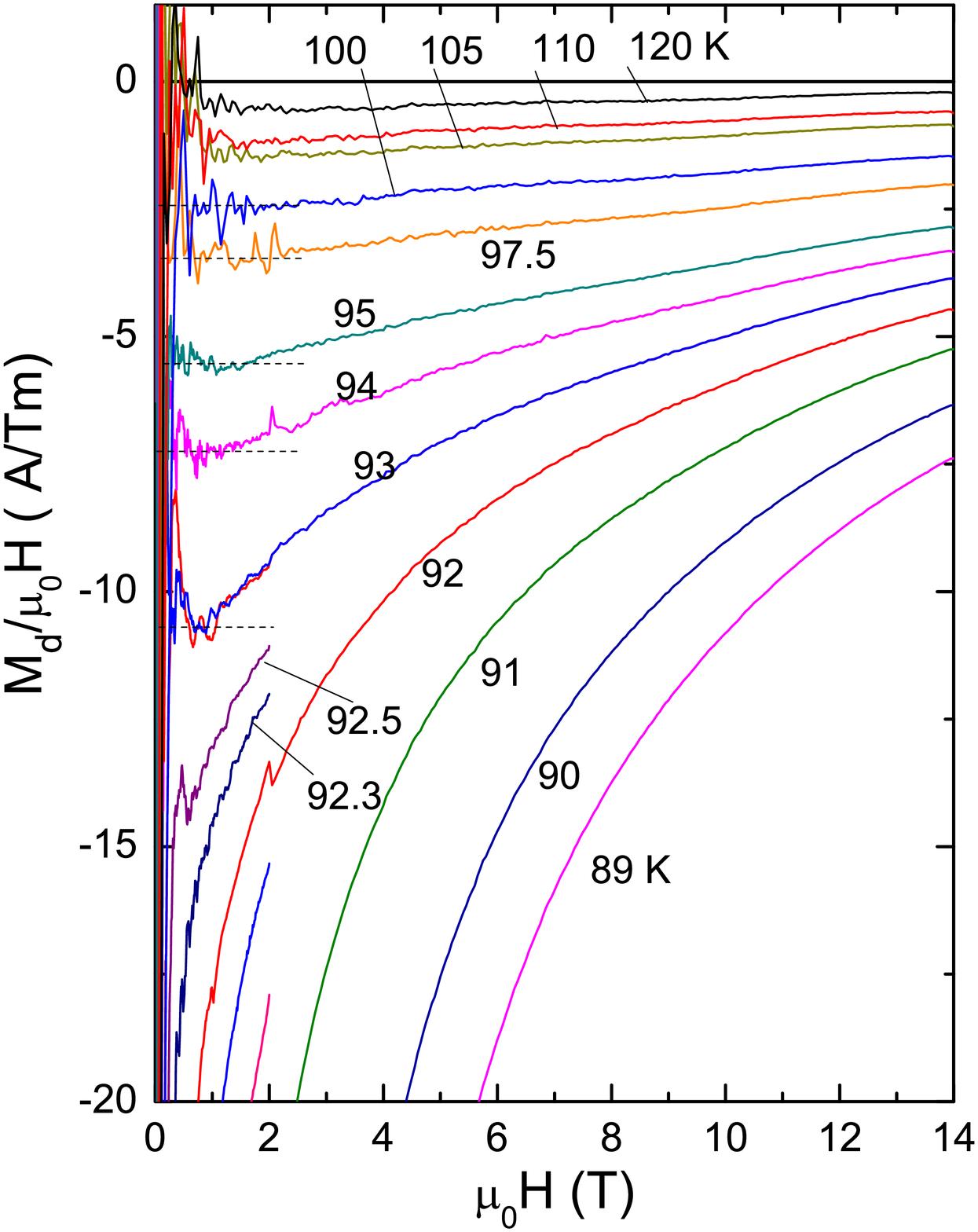}
\caption{\label{figChiH} (color online) 
Plots of the susceptibility $\chi = M/H$ vs. $\mu_0H$ in OPT YBCO at
selected $T$ ($T_c$ = 92.5 K). The dashed horizontal lines indicate the linear-response region
for curves near $T_c$. 
}
\end{figure}

We comment on the fits above $T_c$ reported in Ref. \cite{Rey}.
Figure \ref{figChiH} shows plots of the susceptibility $\chi = M/H$.
Above 100 K, $\chi$ is $H$-independent over a large field region.
However, as $T$ falls below 100 K, the linear-response is confined to progressively smaller
field ranges, as mentioned. Nonetheless, as $T\to T_c^{+}$, the linear-response value of $\chi$
(dashed lines in Fig. \ref{figChiH}) does not diverge. 
Instead, the appearance of the Meissner effect (and its subsequent
suppression by $H$) occur at much lower field scales as discussed. 
Below $T_c$, the hallmark non-monotonicity of
$M_d$ vs. $H$ described above is harder to see in plots of $\chi$ vs. $H$
because of the strong field variation caused by dividing a relatively flat profile by $H$.

\begin{figure}[t]
\includegraphics[width=8 cm]{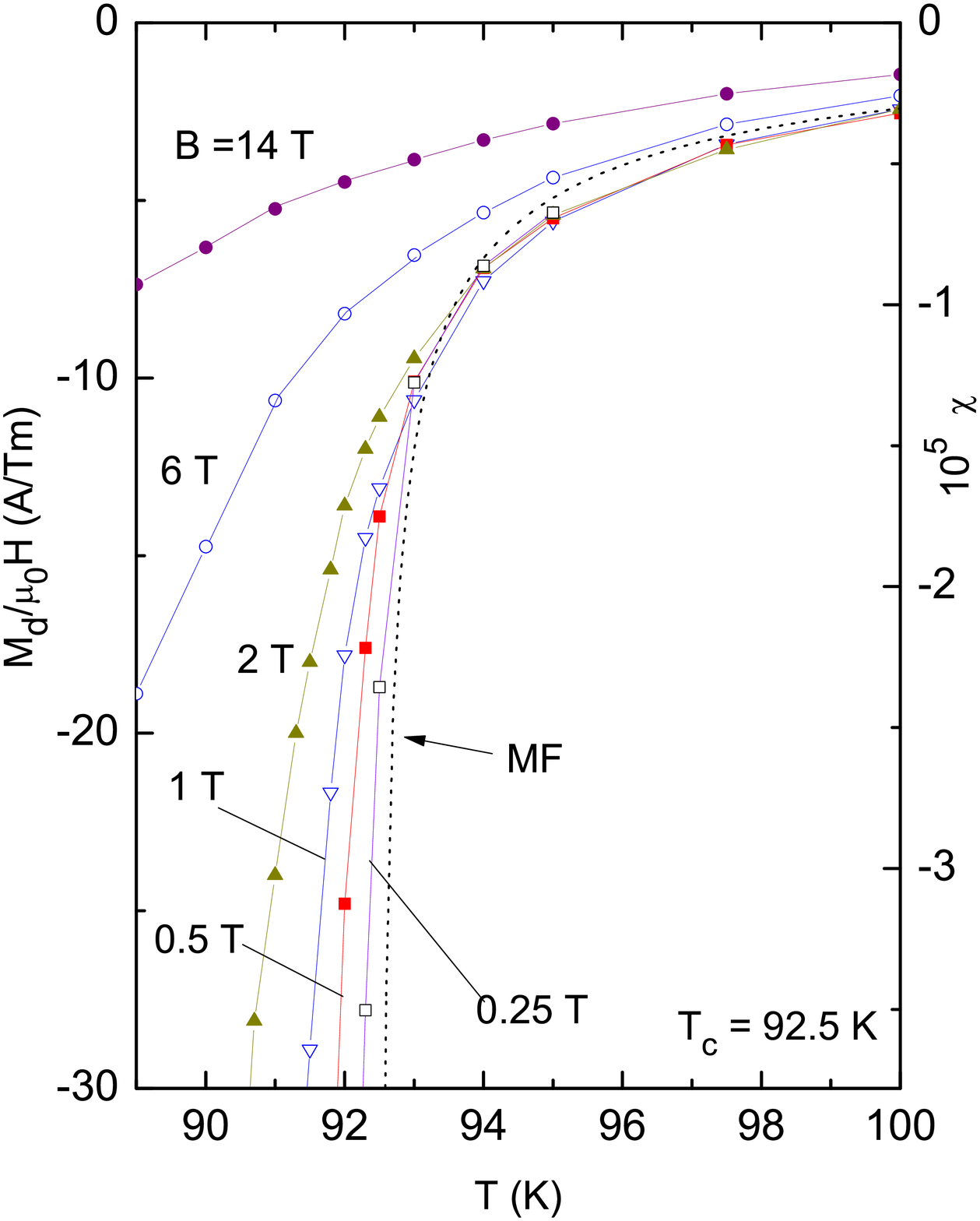}
\caption{\label{figChiT} (color online) 
The $T$ dependence of the susceptibility in OPT YBCO for selected values of $H$.
Above $T_c$ (92.5 K), $\chi$ is $H$ independent below 2 T (linear response region).
Below $T_c$, $\chi$ becomes strongly $H$ dependent at low $H$ because of the
field suppression of phase locking in the vicinity of $H_{cusp}$. At large $H$
(curves at 6 and 14 T), $\chi$ varies smoothly through $T_c$. The dashed line is the
mean-field (MF) fit to Eq. \ref{eq:chi} with $\eta = 3.78\times 10^{-7}$ and $B_{LD} = 0.11$. 
}
\end{figure}

Moving to the fluctuation regime above $T_c$, we show in Fig. \ref{figChiT}
the plots of $\chi$ vs. $T$ at selected values of $H$. Above $T_c$, all values
of $\chi$ below 2 T collapse to a single curve, which defines the linear-response
fluctuation susceptibility that can be directly compared with mean-field theories~\cite{Schmid,Yamaji,Vidal94}. 
For illustration, we show as a dashed curve the best fit to the linear-response MF expression~\cite{Yamaji} 
\be
\chi(T) = -\frac{\eta}{\sqrt{[\epsilon^2+\epsilon B_{LD}] }},
\label{eq:chi}
\ee
with $\eta$ an adjustable numerical factor and $\epsilon = (T-T_c)/T_c$ (the dashed
curve has $\eta = 3.78\times 10^{-7}$). 
For the LD parameter, we used the value proposed by Rey \etal~\cite{Rey} ($B_{LD}$ = 0.11).

While the fit can account for the overall magnitude of $\chi$ above 100 K with reasonable
parameters, as shown by Rey \etal, the functional
form in Eq. \ref{eq:chi} does not describe the data trend all that well. 
Between 94 and 98 K, it underestimates
$|\chi|$ by as much as 15$\%$, and overestimates its value at 92.5 K. A similar pattern
of overshooting and undershooting is also evident in Fig. 1(a) of Rey \etal.

At large $H$ (curves at 6 and 14 T), $\chi$ varies smoothly through $T_c$ without 
detectable change in slope. This reflects the continuity of the diamagnetic response
in strong fields described above. Significantly, at low $H$, 
the linear-response segment of $\chi$ suddenly becomes strongly $H$-dependent 
even in very weak $H$. As we discussed, this reflects the non-monotonicity
that appears above the field $H_{cusp}$. These features cannot be described by
the Gaussian approach of Rey \etal~\cite{Rey}.

\section{Summary}
In attempting to understand the fluctuation signals of the magnetization in OPT YBCO, it is
essential to view the results above and below $T_c$. We show that the non-monotonic curves of
$M_d$ vs. $H$ just below $T_c$ are consistent with the loss of long-range phase coherence in a 
layered superconductor with extremely large pair-binding energy within each layer, but a $c$-axis 
coupling (between bilayers) that is suppressed by a few Teslas close to $T_c$. These features are incompatible with the Gaussian
picture. In addition, when discussing the thermodynamics of hole-doped cuprates, we argue that it is vital to 
recognize the elephant in the room, namely the vortex liquid above and below $T_c$. The dominant presence of the vortex liquid
alters qualitatively the profiles of the $H_{c2}$ curve, pushing its closure to a temperature higher
than $T_c$. As a result, the corner-stone feature of the Gaussian approach, namely an $H_{c2}$ line
that terminates at $T_c$, is absent in OPT YBCO (and other hole-doped cuprates). 
This contradicts a prediction of Ref. \cite{Rey}.

In the phase-disordering picture, the $d$-wave pairing gap $\Delta$ must persist high above $T_c$.
Spectroscopic experiments are increasingly able to distinguish 
between the gap in the vicinity of the node from the much larger antinodal gap.
The persistence of the nodal $\Delta$ above $T_c$ has now been reported in 
scanning tunneling microscopy~\cite{Yazdani} and photoemission experiments~\cite{Dessau,Kaminsky} 
on OPT Bi 2212,
and in $c$-axis infrared reflectivity experiments on YBCO ~\cite{Tajima}.
The vanishing of the Meissner effect above $T_c$ reflects the collapse
of the inter-bilayer phase-coupling. Since the intra-bilayer phase coupling 
is much stronger, one might expect that the concomitant phase
rigidity can be observed experimentally above $T_c$. 
Recently, this was detected by Dubroka \etal~\cite{Bernhard}
as a Josephson plasma resonance that persists above $T_c$ over a broad doping range in YBCO 
(as high as 180 K in the underdoped regime).

We have benefitted from valuable discussions with Zlatko Tesanovic, Oskar Vafek, Steve Kivelson, 
Ashvin Vishwanath, Sri Raghu, Ali Yazdani, Christian Bernhard and P. W. Anderson. Support from the U.S. 
National Science Foundation under MRSEC grant
DMR 0819860 is gratefully acknowledged.

\end{document}